\documentclass[fleqn,12pt,twoside]{article}
\usepackage{espcrc1,epsfig,axodraw}

\newcommand{\ba}{\begin{eqnarray}}
\newcommand{\ea}{\end{eqnarray}}

\newcommand{\fig}{Fig.~}
\newcommand{\figs}{Figs.~}
\newcommand{\eq}{Eq.~}
\newcommand{\eqs}{Eqs.~}
\newcommand{\nr}[1]{(\ref{#1})}

\newcommand{\fr}[2]{{\frac{#1}{#2}\,}}
\newcommand{\msbar}{{\overline{\mbox{\rm MS}}}}
\newcommand{\lambdamsbar}{{\Lambda_{\overline{\rm MS}}}}
\renewcommand{\(}{\left(}
\renewcommand{\)}{\right)}
\newcommand{\lb}{\left\{}
\newcommand{\rb}{\right\}}
\newcommand{\lk}{\left[}
\newcommand{\rk}{\right]}
\newcommand{\ld}{\left.}

\renewcommand{\d}{\delta}

\renewcommand{\l}{\lambda}
\newcommand{\6}{\partial}
\newcommand{\tr}{{\rm Tr}\,}
\newcommand{\sy}[3]{{\textstyle #1\frac{#2}{#3}}}
\newcommand{\rmi}[1]{{\mbox{\scriptsize #1}}}

\newcommand{\pib}[1]{\;\parbox[c]{36pt}{\begin{picture}(22.5,30)(0,0)
 \SetWidth{1.0}\SetScale{1.0} #1 \end{picture}}\;}
\newcommand{\spic}[1]{\;\parbox[c]{21pt}{\begin{picture}(21,21)(0,0)
 \SetWidth{1.0}\SetScale{0.7} #1 \end{picture}}\;}
\newcommand{\spicb}[1]{\;\parbox[c]{32pt}{\begin{picture}(32,21)(0,0)
 \SetWidth{1.0}\SetScale{0.7} #1 \end{picture}}\;}
\newcommand{\spicc}[1]{\;\parbox[c]{42pt}{\begin{picture}(42,21)(0,0)
 \SetWidth{1.0}\SetScale{0.7} #1 \end{picture}}\;}
\def\Asc(#1,#2)(#3,#4,#5){\CArc(#1,#2)(#3,#4,#5)}
\def\Lsc(#1,#2)(#3,#4){\Line(#1,#2)(#3,#4)}
\def\STTopoVRoo(#1){\;\spic{#1(15,15)(15,0,180) #1(15,15)(15,180,360)%
 \GCirc(0,15){5}{0.75} \Text(0,10.5)[c]{$\scriptscriptstyle 1$}
 \GCirc(30,15){5}{0.75} \Text(21,10.5)[c]{$\scriptscriptstyle 1$}}\;}
\def\STTopoVRor(#1){\;\spic{#1(15,15)(15,0,180) #1(15,15)(15,180,360)%
 \GCirc(0,15){5}{0.75} \Text(0,10.5)[c]{$\scriptscriptstyle 1$}
 \GBoxc(30,15)(9,9){0.75} \Text(21,10.5)[c]{$\scriptscriptstyle 2$} }\;}
\def\STTopoVRoi(#1){\;\spic{#1(15,15)(15,0,180) #1(15,15)(15,180,360)%
 \GCirc(0,15){5}{0.75} \Text(0,10.5)[c]{$\scriptscriptstyle 1$} 
 \GCirc(30,15){5}{0.75} \Text(21,10.5)[c]{$\scriptscriptstyle 2$} }\;}
\def\STTopoVRooo(#1){\spic{#1(15,15)(15,-30,90) #1(15,15)(15,90,210)%
 #1(15,15)(15,210,330) 
 \GCirc(15,30){5}{0.75}\Text(10.5,21)[c]{$\scriptscriptstyle 1$}%
 \GCirc(2,7.5){5}{0.75}\Text(1.4,5.25)[c]{$\scriptscriptstyle 1$}%
 \GCirc(28,7.5){5}{0.75}\Text(19.6,5.25)[c]{$\scriptscriptstyle 1$}}}
\def\SToptVE(#1,#2){\spicc{#1(15,15)(15,0,360) #2(45,15)(15,-180,180)}}
\def\SToptVSblobsh(#1,#2,#3){\spic{#1(15,15)(15,0,180) #2(15,15)(15,180,360)%
 #3(30,15)(0,15) \GCirc(30,15){3}{0.5}}\;}
\def\SToprVVblobbsh(#1,#2,#3,#4,#5){\!\!\spicb{#2(26.25,15)(15,256,76)%
 #3(30,30)(15,30) #1(18.75,15)(15,104,284) #4(15,30)(22.5,0)%
 #5(30,30)(22.5,0) \GCirc(29,28){3}{0.5} \GCirc(16,28){3}{0.5}}\!\!}
\def\SToprVBblobsh(#1,#2,#3,#4){\spicb{#1(30,15)(15,-120,120)%
 #2(30,15)(15,120,240)%
 #3(15,15)(15,60,300) #4(15,15)(15,-60,60) \GCirc(22.5,28){3}{0.5}}}
\def\SPropCircProp(#1,#2){\!\pib{#1(5,15)(15,15) #1(25,15)(35,15)
 \GCirc(20,15){5}{0.75} \Text(20,15)[c]{$\scriptstyle #2$}}}
\def\SPropBoxProp(#1,#2){\!\pib{#1(5,15)(15,15) #1(25,15)(35,15)
 \GBoxc(20,15)(9,9){0.75} \Text(20,15)[c]{$\scriptstyle #2$}}}

\title{Long-distance contributions to the QCD pressure\thanks{talk
given at {\em Statistical QCD}, Bielefeld, August 26-30, 2001.}}

\author{York Schr\"oder\address{Department of Physics,
P.O.Box 64, FIN-00014 University of Helsinki, Finland}\thanks{present 
address: Center for Theoretical Physics, MIT, Cambridge, MA 02139, USA.}}

\begin{document}

\maketitle


\begin{abstract}
The QCD pressure is a most fundamental quantity, for which lattice data is
available up to a few times the critical temperature $T_c$. 
Perturbation theory, even at very high temperatures, has serious
convergence problems. Combining analytical and 3d numerical methods,
we show that it is possible to compute the QCD pressure from about
$2 T_c$ to infinity. We also describe an algorithm to 
generate and classify high order Feynman diagrams which is tailored 
to minimize computational effort.
\end{abstract}


\setcounter{footnote}{0}

\mbox{}

\noindent{\bf Introduction.}
The properties of QCD matter are expected to change above a
critical temperature of the order of 200 MeV. 
While the low-temperature phase is governed by bound states,
such as mesons, the high-temperature phase should, due to
asymptotic freedom, look more like a gas of free quarks and gluons.
Any observable witnessing this change is therefore a 
potential candidate for (direct or indirect) measurements in 
heavy-ion collision experiments.

One such observable clearly is the free energy density of the system.
The rough picture is that it is, according to the Stefan-Boltzmann law, 
proportional to the number of effective degrees of freedom. 
For vanishing baryon density $\mu_b=0$ and at temperature $T$,
the free energy density of QCD is simply given by the functional integral
\ba
f &=& 
-\fr{T}V\ln \int{\cal D}[A_\mu^a\bar\psi_f\psi_f]
\exp\( -\int_0^{1/T} \!\!d\tau \int_V \!d^3x\lk
\fr14 F_{\mu\nu}^aF_{\mu\nu}^a
+\sum_f\bar\psi_f\gamma_\mu D_\mu \psi_f \rk \) \;.
\ea
Note that, in the thermodynamic limit of infinite volume $V$, 
the pressure $p$ of the plasma is given directly by $p=-f$. 
Below, we will choose to display results for the pressure. 

The most direct way to evaluate this integral would now
be to measure it numerically on the lattice. In fact, this
has been done by a number of groups. While the results for
$N_f=0$ are rather complete, they are rapidly developing for 
a finite number of fermion flavours $N_f$ (e.g. \cite{lat,latNf}). 
The general picture emerging from these
lattice simulations is the following: Normalizing the pressure 
to zero below $T_c$ (in practice, one can only measure derivatives
of the free energy, leaving an integration constant to be fixed), 
it rises sharply in the interval $(1-2)T_c$, to level off at a few times
$T_c$. At the highest temperatures used in the simulations,
typically $(4-5)T_c$,
the deviation to the Stefan-Boltzmann limit is of the order of $15\%$.
This general picture is surprisingly stable with respect to 
different values of $N_f$.
At even higher temperatures, the pressure is then expected to
asymptotically approach the ideal-gas limit
$p_0(T)=(\pi^2 T^4/45)(N_c^2\!-\!1+(7/4)N_cN_f)$, where $N_c$ denotes 
the number of colours. 

\begin{figure}[thb]
\begin{minipage}[t]{75mm}
\epsfig{file=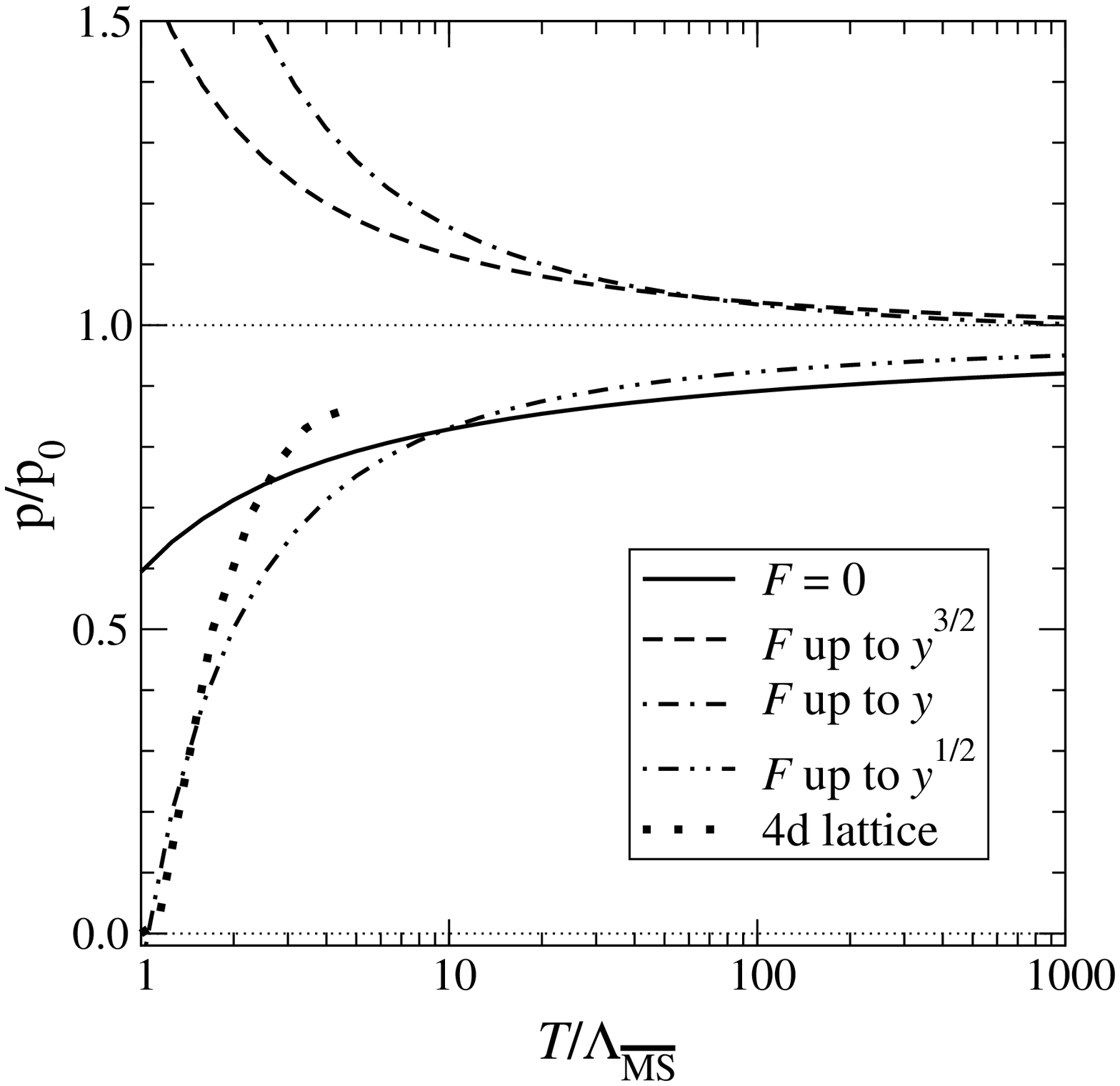,width=74mm}
\vskip -0.9truecm
\caption{Perturbative pressure vs. 
lattice data for $N_f=0$ (from \cite{fqcd}).
The notation is like in \eq\nr{newNot}, corresponding to $p$ up to $g^2$, 
$g^3$, $g^4$ and $g^5$.}
\vskip -0.6truecm
\label{fig:pert}
\end{minipage}
\hspace{\fill}
\begin{minipage}[t]{75mm}
\epsfig{file=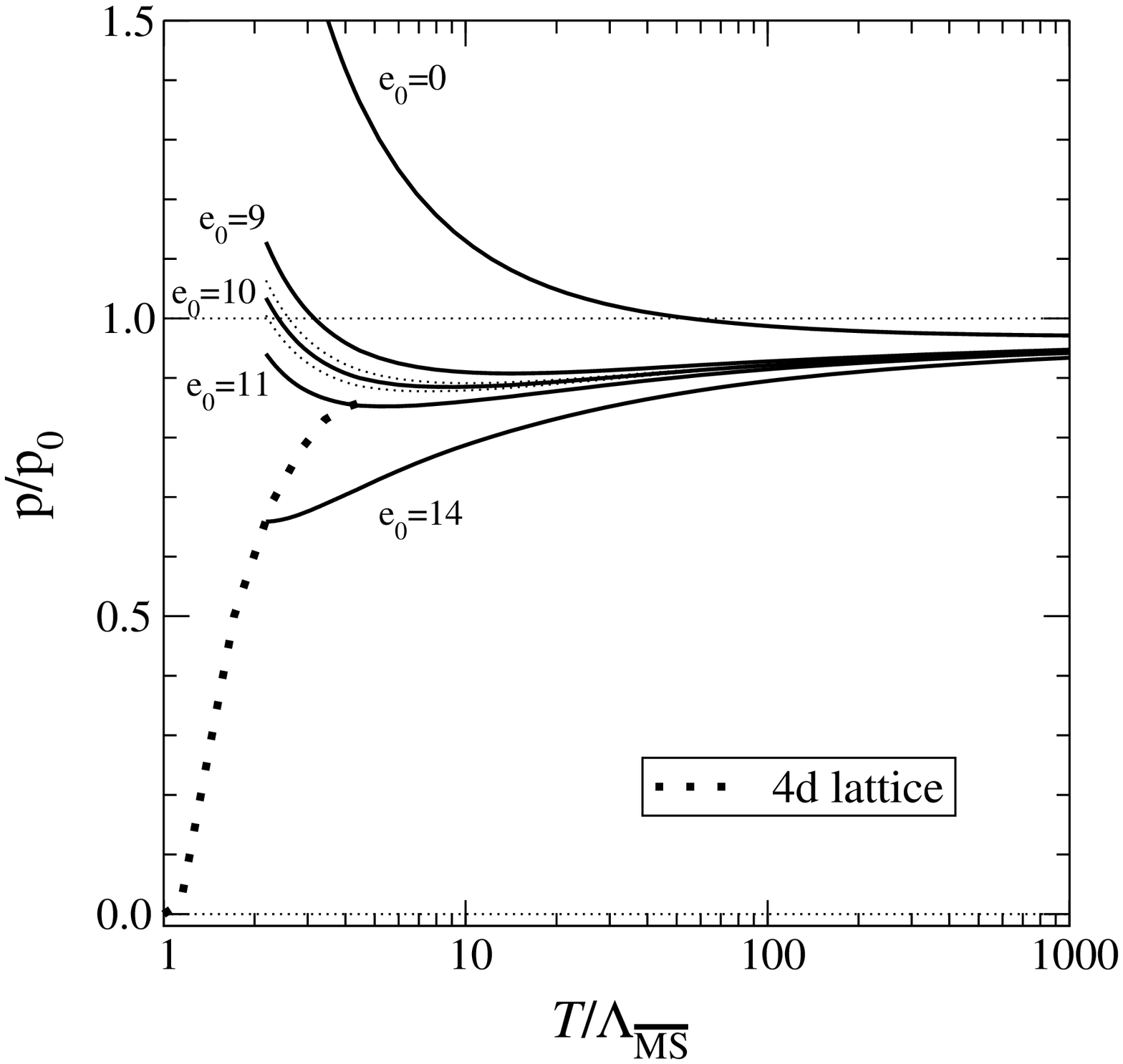,width=74mm}
\vskip -0.9truecm
\caption{The pressure after inclusion of the long-distance part 
according to \eq\nr{3dF} (from \cite{fqcd}). 
Statistical errors are shown only for $e_0=10$.}
\vskip -0.6truecm
\label{fig:press}
\end{minipage}
\end{figure}

It turns out that this deviation of $15\%$ is too large to be understood 
in terms of ordinary perturbation theory. 
In a series of impressive works, the expansion has been driven to 5th order 
in the gauge coupling $g$ \cite{zk}, 
\ba \label{pert4d}
\fr{p(T)}{p_0(T)}&=&
1+c_2g^2+c_3g^3+(c^{\prime}_4\ln g+c_4)g^4+c_5g^5+{\cal O}(g^6\ln g,g^6)
\;.
\ea   
While the coefficients are known analytically, convergence properties 
are extremely poor, cf.~\fig\ref{fig:pert}, at least for all physically 
relevant temperatures.

Facing the poor convergence of the perturbative series, in the past
few years a lot of effort has gone into refined and/or alternative 
approaches, in order to gain an analytic understanding of the 
high-temperature behaviour of the pressure. 
The spectrum ranging from Pad\'{e}-Borel resummations over using effective
masses to employing selfconsistent approximations \cite{bir}, a general 
feature of these works is the suppression of infrared (long-distance) 
effects. While this suppression does not seem to be crucial in the 
computation of the pressure (which appears to be a short-distance 
dominated observable)\footnote{In fact, the last method mentioned 
connects to the lattice data available quite nicely from above.}, 
the aim of this talk is to review our 
framework to resum the long-distance contributions to the 
pressure to all orders \cite{fqcd}. 


\mbox{}

\noindent{\bf Combined analytic and 3d numerical method.}
%
A way to understand the poor convergence of the ordinary perturbative
expansion is the observation that at small gauge coupling $g$,
the system undergoes dimensional reduction (see e.g. \cite{bn} and
references therein). 
The scale hierarchy $gT\ll\pi T$
allows to perturbatively construct an effective theory for the 
``soft'' modes (momenta $\propto gT$) by integrating out the
``hard'' modes ($\propto \pi T$). In the case of QCD, 
this effective theory is a
3d SU($N_c$) + adjoint Higgs model:
\ba \label{adjH}
{\cal L}_{\rm 3d} &=& 
\fr14 \tr F_{ij}^2 +\fr12 \tr [D_i,A_0]^2 +\fr12m_D^2 \tr A_0^2 
+\fr14\lambda_A (\tr A_0^2)^2 +\d{\cal L}_{\rm 3d} \;.
\ea
While the last term represents higher-order operators, which we do
not take into account here
because their contributions are parametrically suppressed \cite{ad}, 
the parameters of the first terms 
($g_3$, $m_D^2$, $\l_A$) are related to the physical 
parameters of the full 4d theory ($T$, $\lambdamsbar$).
Using 
optimized next-to-leading order perturbation theory \cite{ad} 
(let us introduce dimensionless parameters\footnote{The 3d gauge 
coupling $g_3^2$ has the dimension of a mass.} $x$, $y$ and
set $N_f=0$, $N_c=3$ here), they read:
\ba
\frac{g_3^2}{T} = \fr{8\pi^2/11}{\ln(6.742 T/\lambdamsbar)}
\!\!\quad,\quad\!\!
x = \fr{\l_A}{g_3^2} = \fr{3/11}{\ln(5.371 T/\lambdamsbar)} 
\!\!\quad,\quad\!\!
y = \fr{m_D^2}{g_3^4} = \fr3{8\pi^2 x}\!+\!\fr9{16\pi^2} \;.
\ea

Two comments are now in order. First, ${\cal L}_{\rm 3d}$ can be used to 
reproduce \eq\nr{pert4d} in a technically simple way \cite{bn}. 
This shows that the 
bad convergence is due precisely to the soft degrees of freedom,
and it provides a clearer understanding of the contributions from
separate physical scales. 

Second, the effective theory is confining,
hence non-perturbative \cite{nonpert}. 
This fact directly leads us to the conclusion
that the only way to systematically include the long-distance 
contributions to the pressure is to treat ${\cal L}_{\rm 3d}$ on the
lattice.

To proceed, we rewrite the pressure, up to hard-scale $g^6$ contributions, 
as
\ba \label{newNot}
\fr{p(T)}{p_0(T)} &=& 1 -\fr{5x}2 -\fr{45}{8\pi^2} \(\fr{g_3^2}T\)^3 
\lk {\cal F}_{\overline{\rm MS}}(x,y) -\fr{24y}{(4\pi)^2} 
\(\ln\fr{\bar\mu_{\rm {3d}}}T +\d\) \rk \;,
\ea
where $\d\sim10^{-4}$ and the dependence on the scale $\bar\mu_{\rm {3d}}$, 
which originates from an infrared divergence of the 4d part, cancels
against a similar ultraviolet term\footnote{This is precisely 
the way the effective theory is set up: dependence on a matching scale
has to cancel.} 
in the dimensionless 3d free energy density 
${\cal F}_{\overline{\rm MS}}(x,y)$,
\ba
{\cal F}_{\overline{\rm MS}} = 
-\frac{1}{Vg_3^6} \ln \int \!\! {\cal D}A 
\exp\(\!-\!\!\int\!\! d^3x {\cal L}_{\rm 3d}\) 
= {\cal F}_{\overline{\rm MS}}(x_0,y_0) 
+\int_{y_0}^{y} \! dy \( 
\fr{\partial {\cal F}_{\overline{\rm MS}}}{\partial y}\! + \fr{d x}{dy} 
\fr{\partial {\cal F}_{\overline{\rm MS}}}{\partial x}
\) \;, \label{3dF}
\ea
which should be measured on the lattice. This requires a measurement of
the quadratic as well as quartic adjoint Higgs field condensates 
(which determine the partial derivatives under the integral), as
well as a (4-loop) perturbative computation in lattice regularization, to match
to the $\msbar$ scheme. We choose to fix the integration
constant perturbatively at high temperatures $T\sim 10^{11} T_c$, 
where one is confident that the expansion converges\footnote{
$T_c$ is of the order of $\lambdamsbar$, the coefficient can be
measured on the lattice.}.


Before displaying first
results for the pressure obtained by the above strategy, let us 
illuminate one aspect in more detail here, namely the setup of 
its high-order perturbative expansion in an efficient manner.


\mbox{}

\noindent{\bf Diagrammatics.}
The enumeration of Feynman diagrams contributing to a specific loop
order together with a derivation of the accompanying symmetry factors
seems to be the 'trivial' part of any perturbative calculation. 
At higher orders, it is our experience that an efficient 
algorithmic setup of this initial step proves necessary. 
This does not only 
assure completeness of the required set of diagrams, but it also
has the potential of 
streamlining the subsequent integration step considerably by grouping
together sets of related diagrams, thereby avoiding an unnecessary
repetition of subdiagram computations \cite{sd}. 

The main idea is to utilize the very efficient notion of skeleton 
(2-particle-irreducible, 2PI) diagrams to achieve the above-mentioned 
grouping (into {\em skeleton} and {\em ring} diagrams, see below). 
For simplicity, we will turn to a 
generic bosonic $\phi^3+\phi^4$ theory here.
The skeleton expansion for the free energy as a functional of the
full propagator $D$ reads \cite{LW}
\ba \label{Fskel}
-F[D] = -\fr12\(\tr\ln D^{-1} +\tr\Pi[D] D\) +\Phi[D] \;.
\ea 
Here $\Phi[D]$ collects all 2PI vacuum diagrams.
The full propagators $D$ 
are related to their corresponding self-energies
by $D^{-1}=\Delta^{-1}-\Pi$ 
where $\Delta$ are the free propagators. 
The partition function has an 
extremal property, such that the variation
of $F$ with respect to the full propagator
vanishes,
giving a relation between skeletons and self-energies,
$\6_D\, \Phi[D] \;=\; \fr12\,\Pi[D]$.
Pictorially, this corresponds to obtaining 
a self-energy by ``cutting a propagator'' 
in all possible ways in the set of vacuum skeletons.
Hence, knowing the skeletons alone provides full information.

To utilize \eq\nr{Fskel}, we need to rewrite it as a loop expansion,
which can be achieved by a straightforward Taylor expansion around
the free propagator $\Delta$. The result, up to 4-loop order, reads
diagrammatically (denoting by $F_0$ the non-interacting result) 
\ba \label{graphic}
-F = -F_0 +\Phi_2[\Delta] 
+\( \Phi_3[\Delta] \sy+14\STTopoVRoo(\Asc) \) 
+\( \Phi_4[\Delta] \sy+16\STTopoVRooo(\Asc) 
\sy+12\STTopoVRoi(\Asc) \sy+14\STTopoVRor(\Asc) \) \;,
\ea
clarifying our terminology of having achieved a separation into
{\em skeleton} and {\em ring} diagrams. 
The notation here is that $\Phi_n[\Delta]$ are $n$\/-loop skeletons
built up of free propagators, while
a circle/square with $n$ inside denotes what we term an $n$\/-loop 
irreducible/reducible self-energy $\Pi^\rmi{irr}_n$/$\Pi^{\rmi{red}}_n$. 
More precisely, the irreducible self-energies derive directly from
the skeletons, 
while reducible self-energies have (at least\footnote{In higher orders,
reducible self-energies divide naturally into classes according to the
number and type of self-energy insertions. 
At 4-loop level there is only one class, so we do not introduce
extra indices here.}) 
one further self-energy insertion, 
\ba
\SPropCircProp(\Lsc,n) = \Pi_n^\rmi{irr}[\Delta]=2\6_\Delta\Phi_{n+1}[\Delta]
\quad,\;\,
\SPropBoxProp(\Lsc,n) = \Pi_n^\rmi{red}[\Delta]=\ld\Pi_{n-1}^\rmi{irr}[\Delta
+\Delta\Pi_1^\rmi{irr}\Delta]\right|_{n{\rm-loop}}\;.
\ea
Hence it is now clear that \eq\nr{graphic} 
expresses the loop expansion of the free energy in an economic way 
in terms of the irreducible skeletons $\Phi_n[\Delta]$:
either as direct contributions, or as self-energy insertions, 
obtained from the same skeletons by derivatives.

While everything is reduced to knowing the skeletons and their
symmetry factors now, the key advance is that one can write down
a closed exact equation which directly generates the n-loop skeletons
\cite{sd}:
\ba \label{SDvac} 
\Phi_n[\Delta] = \fr1{n-1} \lb 
\sy{}1{12} \SToptVSblobsh(\Asc,\Asc,\Lsc)  
\sy+18 \SToptVE(\Asc,\Asc)
\sy+18 \SToprVVblobbsh(\Asc,\Asc,\Lsc,\Lsc,\Lsc)
\sy+1{24} \SToprVBblobsh(\Asc,\Asc,\Asc,\Asc)
\rb_{n}, \quad n\ge 2\;.
\ea
Lines are bare propagators $\Delta$, and vertices are bare and
full ones (shaded blob), respectively. The full vertices are 
generated in the standard way by the tower of Schwinger-Dyson equations, 
which we do not repeat here. For a functional derivation as well as
a graphical representation see, e.g., \cite{Cvi}. 

It is now straightforward to utilize the concept of \eqs\nr{graphic},\nr{SDvac}
to generate all diagrams needed for the theory in \eq\nr{adjH}, to serve 
as a starting point for the (up to) 4-loop computations needed. 


\mbox{}

\noindent{\bf Discussion.}
As a first result obtained along the strategy outlined 
above,
\fig\ref{fig:press} shows the normalized pressure. The integration constant
has been fixed perturbatively on the 3-loop level, allowing for 
an additional constant $e_0$, which represents an (up to now) 
unknown $g_3^6$ contribution. In principle, this constant can be
determined in a setup equivalent to the above, after splitting
off its perturbative part: A further reduction
step relates $e_0$ to the free energy of 3d pure gauge theory, which
is at the core of the famous non-perturbative $g^6$ term, but can
nevertheless be determined on the lattice.
On the lattice side, we have only included the 
quadratic scalar condensate, while at temperatures closer to $T_c$,
the quartic one will become important as well. 

While more work is required (and in progress), 
we wish to point out two important trends seen in 
\figs\ref{fig:pert},\ref{fig:press}: 
First, the outcome is sensitive to the non-perturbative 
parameter $e_0$, which in principle can be determined by additional
computations. Clearly, there exists a range for that parameter 
(say, ${\cal O}(10)$),
which leads to a sensible result. 

Second, at $T>30\lambdamsbar$ the curves for $g^5$ ($y^{\fr12}$) 
and $e_0=10$ fall
almost on top of each other, signalling a cancellation of all higher-order 
terms (determined here by the quadratic Higgs condensate) against the large 
non-perturbative $g^6$ contribution. Hence, in this temperature range the
pressure is indeed dominated by short-distance effects.

While we have mostly presented results for QCD at
$N_f=0$ and zero baryon number, 
inclusion of $N_f$ fermion flavours as well as a baryon chemical
potential $\mu_b$ pose no further complications, and hence provide 
for a natural extension of this investigation.



\mbox{}

\noindent{\bf Acknowledgements.}
This work was supported in part by the EU TMR network 
ERBFMRX-CT-970122. 
It is a pleasure to thank K.~Kajantie,
M.~Laine and K.~Rummukainen for an enjoyable collaboration on the 
matter presented in the above.


\end{document}